\documentclass[twocolumn]{aastex63}
\usepackage{amsmath}

\newcommand{\nustar}{\textit{NuSTAR}}

\shorttitle{{\it NuSTAR} Discovery of a Compton-thick DOG}
\shortauthors{Toba et al.}

\begin{document}

\title{\boldmath$NuSTAR$ Discovery of a Compton-thick Dust-obscured Galaxy WISE J0825+3002}

\correspondingauthor{Yoshiki Toba}
\email{toba@kusastro.kyoto-u.ac.jp}

\author[0000-0002-3531-7863]{Yoshiki Toba}
\affiliation{Department of Astronomy, Kyoto University, Kitashirakawa-Oiwake-cho, Sakyo-ku, Kyoto 606-8502, Japan}
\affiliation{Academia Sinica Institute of Astronomy and Astrophysics, 11F of Astronomy-Mathematics Building, AS/NTU, No.1, Section 4, Roosevelt Road, Taipei 10617, Taiwan}
\affiliation{Research Center for Space and Cosmic Evolution, Ehime University, 2-5 Bunkyo-cho, Matsuyama, Ehime 790-8577, Japan}

\author[0000-0002-9754-3081]{Satoshi Yamada}
\affiliation{Department of Astronomy, Kyoto University, Kitashirakawa-Oiwake-cho, Sakyo-ku, Kyoto 606-8502, Japan}

\author[0000-0001-7821-6715]{Yoshihiro Ueda}
\affiliation{Department of Astronomy, Kyoto University, Kitashirakawa-Oiwake-cho, Sakyo-ku, Kyoto 606-8502, Japan}

\author[0000-0001-5231-2645]{Claudio Ricci}
\affiliation{N\'ucleo de Astronom\'{\i}a de la Facultad de Ingenier\'{\i}a, Universidad Diego Portales, Av. Ej\'ercito Libertador 441, Santiago, Chile}
\affiliation{Kavli Institute for Astronomy and Astrophysics, Peking University, Beijing 100871, People's Republic of China}

\author[0000-0003-1780-5481]{Yuichi Terashima}
\affiliation{Graduate School of Science and Engineering, Ehime University, 2-5 Bunkyo-cho, Matsuyama, Ehime 790-8577, Japan}
\affiliation{Research Center for Space and Cosmic Evolution, Ehime University, 2-5 Bunkyo-cho, Matsuyama, Ehime 790-8577, Japan}

\author[0000-0002-7402-5441]{Tohru Nagao}
\affiliation{Research Center for Space and Cosmic Evolution, Ehime University, 2-5 Bunkyo-cho, Matsuyama, Ehime 790-8577, Japan}

\author[0000-0003-2588-1265]{Wei-Hao Wang}
\affiliation{Academia Sinica Institute of Astronomy and Astrophysics, 11F of Astronomy-Mathematics Building, AS/NTU, No.1, Section 4, Roosevelt Road, Taipei 10617, Taiwan}

\author[0000-0002-0114-5581]{Atsushi Tanimoto}
\affiliation{Department of Astronomy, Kyoto University, Kitashirakawa-Oiwake-cho, Sakyo-ku, Kyoto 606-8502, Japan}

\author[0000-0002-6808-2052]{Taiki Kawamuro}
\affiliation{National Astronomical Observatory of Japan, 2-21-1 Osawa, Mitaka, Tokyo 181-8588, Japan}











\begin{abstract}

We report the discovery of a Compton-thick (CT) dust-obscured galaxy (DOG) at $z$ = 0.89, WISE J082501.48+300257.2 (WISE0825+3002), observed by {\it Nuclear Spectroscopic Telescope Array} ({\it NuSTAR}).
X-ray analysis with the XCLUMPY model revealed that hard X-ray luminosity in the rest-frame 2--10 keV band of WISE0825+3002 is $L_{\rm X}$ (2--10 keV) = $4.2^{+2.8}_{-1.6} \times 10^{44}$ erg s$^{-1}$ while its hydrogen column density is $N_{\rm H}$ = $1.0^{+0.8}_{-0.4} \times 10^{24}$ cm$^{-2}$, indicating that WISE0825+3002 is a mildly CT active galactic nucleus (AGN). 
We performed the spectral energy distribution (SED) fitting with {\sf CIGALE} to derive its stellar mass, star formation rate, and infrared luminosity.
The estimated Eddington ratio based on stellar mass and integration of the best-fit SED of AGN component is $\lambda_{\rm Edd}$ = 0.70, which suggests that WISE0825+3002 harbors an actively growing black hole behind a large amount of gas and dust.
We found that the relationship between luminosity ratio of X-ray and 6 $\micron$, and Eddington ratio follows an empirical relation for AGNs reported by \cite{Toba_19a}.

\end{abstract}



\section{Introduction}
\label{intro}

In the last two decades, it has been revealed that almost all galaxies harbor a supermassive black hole (SMBH) with a mass of $10^{5-10} M_{\sun}$ in their centers. 
The BH masses are well-correlated with those of the spheroid component of their host galaxies, suggesting that SMBHs and their host galaxies coevolve \citep[e.g.,][]{Magorrian,Marconi,Kormendy}.
The physics of the co-evolution of galaxies and SMBHs has not been constrained observationally  although this is the subject of intense theoretical investigation \citep[e.g.,][]{Hopkins}.
This is because many previous studies are based on optically selected samples, which did not go deep enough to find heavily obscured active galactic nuclei (AGNs), for example, Compton-thick (CT) AGNs with line-of-sight hydrogen column densities of $N_{\rm H} \gtrsim 1.5 \times 10^{24}$ cm$^{-2}$ \citep[e.g.,][]{Ricci_15,Koss}.
In the context of BH growth through a major merger, recent hydrodynamic simulations and observations reported that AGNs with the highest accretion rate are expected to be surrounded by a large amount of gas and dust \cite[e.g.,][]{Narayanan,Ricci_17,Blecha,Yamada2019}.
For a full understanding of the physics of galaxy--SMBH co-evolution, it is crucial to search for actively accreting galaxy--SMBH systems including CT--AGNs.

In this work, we focus on infrared (IR)-bright dust-obscured galaxies (DOGs) \citep{Toba_15,Toba_17,Noboriguchi}
as a key population to address this issue.
The definition of IR-bright DOGs is (i) $i - [22] > 7.0$ in AB magnitude, where $i$ and [22] are $i$-band and 22 $\micron$ magnitude, respectively and (ii) flux density at 22 $\micron$ $>$ 1 mJy that is typically an order of magnitude brighter than that of previously discovered IR-faint DOGs \citep{Dey,Fiore_08}.
\cite{Toba_16} have performed a systematic search for IR-bright DOGs by using the Sloan Digital Sky Survey \citep[SDSS; ][]{York} Data Release 12 \citep{Alam} and the {\it Wide-field Infrared Survey Explorer} \citep[{\it WISE}; ][]{Wright} ALLWISE catalog \citep{Cutri}, and discovered 5311 IR-bright DOGs.
However, the accretion properties of DOGs are poorly understood observationally.
This is partly because the SMBH in DOGs is often highly obscured up to CT level \citep[e.g.,][]{Fiore,Lanzuisi,Georgantopoulos,Corral}, and thus high-sensitivity hard X-ray observations are necessary to constrain the accretion properties in such a dusty population.

In this paper, we present  follow-up observation with the {\it Nuclear Spectroscopic Telescope Array} \citep[{\it NuSTAR}:][]{Harrison} for a candidate of CT--AGN, WISE J082501.48+300257.2 (hereafter WISE0825+3002) at $z = 0.89$ that is drawn from IR-bright DOG sample in \cite{Toba_16}.
The excellent penetrating power of {\it NuSTAR} enables us to unveil the BH properties of WISE0825+3002.
We also perform the spectral energy distribution (SED) analysis to derive its host properties such as stellar mass and star formation rate (SFR).
Throughout this paper, the adopted cosmology is a flat universe with $H_{\rm 0} = 70$ km s$^{-1}$ Mpc$^{-1}$, $\Omega_{\rm M} = 0.27$, and $\Omega_{\rm \Lambda} = 0.73$.

\section{Data and analysis}
\label{data}

\subsection{A candidate of Compton-thick AGN: WISE0825+3002}
WISE0825+3002, a CT--AGN candidate is selected from IR-bright DOG sample in \cite{Toba_16}. 
This sources is also included in the XMM/SDSS serendipitous X-ray survey\footnote{\url{http://members.noa.gr/age/xmmsdss.html}} catalog \citep{Georgakakis}, and its redshift was photometrically estimated to be $z_{\rm photo}$ = 0.89 $\pm$ 0.18 based on the neural network technique \citep[see][for details]{Oyaizu}.
The basic information and the measured flux densities of WISE0825+3002 are summarized in Table \ref{Table}.
Its flux density at 22 $\micron$ is 16.3 mJy (i.e., this object is an extremely IR-bright DOG) and the shape of its mid-IR (MIR) SED can be explained by power-law, which indicates the presence of an AGN (see Section \ref{S_SED} for more quantitative information).
This object is detected by the Very Large Array (VLA) Faint Images of the Radio Sky at Twenty-Centimeters survey \citep[FIRST;][]{Becker,Helfand}.
The rest-frame 1.4 GHz luminosity of WISE0825+3002 is $1.80 \times 10^{25}$ W Hz$^{-1}$ assuming a typical spectral index of radio AGNs, $\alpha_{\rm radio} = 0.7$ \citep[e.g.,][]{Condon}.
Because radio sources with $L_{\rm 1.4\,GHz} > 10^{25}$ W Hz$^{-1}$ are expected to be AGNs \citep{Mauch,Tadhunter}, WISE0825+3002 is an AGN-dominated object.
\cite{Toba_16} conducted an SED analysis with a SED fitting code SEd Analysis using BAyesian Statistics \citep[{\sf SEABASs};][]{Rovilos} \citep[see also][]{Toba_17d,Toba_18}.
The observed MIR data are well-explained by an AGN template with $N_{\rm H}\sim 10^{24}$ cm$^{-2}$.
\cite{Rovilos} also reported that WISE0825+3002 is a CT-AGN candidate based on luminosity ratio of MIR and X-ray.
Therefore, WISE0825+3002 is a good candidate of CT--AGN, and \nustar\ sheds light on BH properties of WISE0825+3002 even if this object is CT \citep[see e.g.,][and references therein]{Marchesi}.

\begin{table}[h]
\begin{footnotesize}
\renewcommand{\thetable}{\arabic{table}}
\centering
\caption{Observed properties of WISE0825+3002.}
\label{Table}
\begin{tabular}{lr}
\tablewidth{0pt}
\hline
\hline
WISE J082501.48+300257.2		&								\\
\hline
R.A. (SDSS) [J2000.0] 			& 	08:25:01.48					\\
Decl. (SDSS) [J2000.0]			& 	+30:02:57.19 				\\
Redshift \citep{Oyaizu}		&	0.89 $\pm$ 0.18			\\
{\it GALEX} NUV  [$\mu$Jy]	&	5.68 	$\pm$ 	1.53			\\
SDSS $u$-band [$\mu$Jy]		&	9.87	$\pm$	1.37			\\
SDSS $g$-band [$\mu$Jy]		&	13.45	$\pm$	0.75			\\
SDSS $r$-band [$\mu$Jy]		&	15.66	$\pm$	0.85			\\
SDSS $i$-band [$\mu$Jy]		&	22.28	$\pm$	1.19			\\
SDSS $z$-band [$\mu$Jy]		&	33.83	$\pm$	4.18			\\
{\it WISE} 3.4 $\micron$ [mJy]	&	0.16 $\pm$ 0.01				\\
{\it WISE} 4.6 $\micron$ [mJy]	&	0.36 $\pm$ 0.02				\\
{\it WISE} 12  $\micron$ [mJy]	&	3.03 $\pm$ 0.18				\\
{\it WISE} 22  $\micron$ [mJy]	&	16.28 $\pm$ 1.06			\\
FIRST 1.4 GHz [mJy]				&	5.39 $\pm$ 0.13				\\
\hline
X-ray spectral analysis	  (Section \ref{S_X}) &								\\
\hline
$L_{\rm X}$ (2-10 keV)  [erg s$^{-1}$]			& 	$4.2^{+2.8}_{-1.6} \times 10^{44}$		\\ 
$N_{\rm H}$ [cm$^{-2}$]							&	$1.0^{+0.8}_{-0.4} \times 10^{24}$		\\
\hline
SED fitting with {\sf CIGALE} (Section \ref{S_SED}) &								\\
\hline
$E(B-V)_{*}$									&	0.21 $\pm$ 0.01					\\
$M_*$ [$M_{\sun}$]								&	$(5.3 \pm 4.4) \times 10^{10}$\\
SFR [$M_{\sun}$ yr$^{-1}$]						&	$(8.5 \pm 3.9) \times 10$	\\
$L_{\rm IR}$ (8-1000 $\micron$) [erg s$^{-1}$]	&	$(1.1 \pm 0.6) \times 10^{46}$\\
$\nu L_\nu$ (6 $\micron$) [erg s$^{-1}$]		&	$(3.4 \pm 1.9) \times 10^{45}$\\
\hline
BH properties	  (Section \ref{S_BH}) &								\\
\hline
$M_{\rm BH}$ [$M_{\sun}$]						& 	$2.5 \times 10^{8}$			\\ 
$\lambda_{\rm Edd}$								&	0.70 \\

\hline
\end{tabular}
\end{footnotesize}
\end{table}

We note that WISE0825+3002 does not satisfy selection criteria of hot DOGs \citep{Eisenhardt,Wu} that are very faint or undetected by {\it WISE} at 3.4 and 4.6 $\micron$, and thus this work may be complemental to previous works based on  {\it NuSTAR} observations of hot DOGs \citep[e.g.,][]{Stern,Assef,Ricci,Vito}.

\subsection{NuSTAR}
{\it NuSTAR} \citep{Harrison} is the first focusing X-ray telescope in orbit that is sensitive to the 3--79 keV band.
It consists of two focal-plane modules (FPMA and FPMB), which offer a 12$\arcmin$~$\times$~12$\arcmin$ field of view (FOV).
{\it NuSTAR} achieves an angular resolution of 18$\arcsec$ full-width at half-maximum (FWHM) with a half-power diameter of 58$\arcsec$.

WISE0825+3002 was observed twice by {\it NuSTAR} (PI Y. Toba) for a net exposure of 6.7~ks on 2018 October 18 (ObsID 60401012002) and for 84.6~ks on 2019 April 16 (ObsID 60401012004). 
The exposure of the first observation was short because it was interrupted by a ToO observation.
The data were processed by using the {\it NuSTAR} data analysis software \textsc{nustardas} v1.8.0 available in
\textsc{heasoft} v6.25 and CALDB released on 2019 May 13. 
The \textsc{nupipeline} script was used to produce calibrated and cleaned event files (with \textsc{saamode=optimized} and \textsc{tentacle=yes}; e.g., \citealt{Iwasawa2017}). 
The source spectra and light curves were extracted with the \textsc{nuproducts} task. 
Photon events were accumulated within a circular region of 30$\arcsec$ radius centered on the peak of the emission in the 3--24 keV band\footnote{We have confirmed that there is no source besides the target around the source region, and the X-ray position of the nucleus matches the optical one with a possible uncertainty in the absolute astrometry of {\it NuSTAR} (8$\arcsec$) \citep[90\% confidence;][]{Harrison}.}, and the background was taken from a source-free
annular region around the source with inner and outer radii of 90$\arcsec$ and 150$\arcsec$, respectively. 
We have confirmed that the spectra and light curves obtained from FPMA and FPMB were consistent with each other. 
We then combined them to increase the photon statistics, using the \textsc{addascaspec} and \textsc{lcmath} tasks,
respectively.

\begin{figure}
    \centering
    \includegraphics[keepaspectratio,scale=0.40]
    {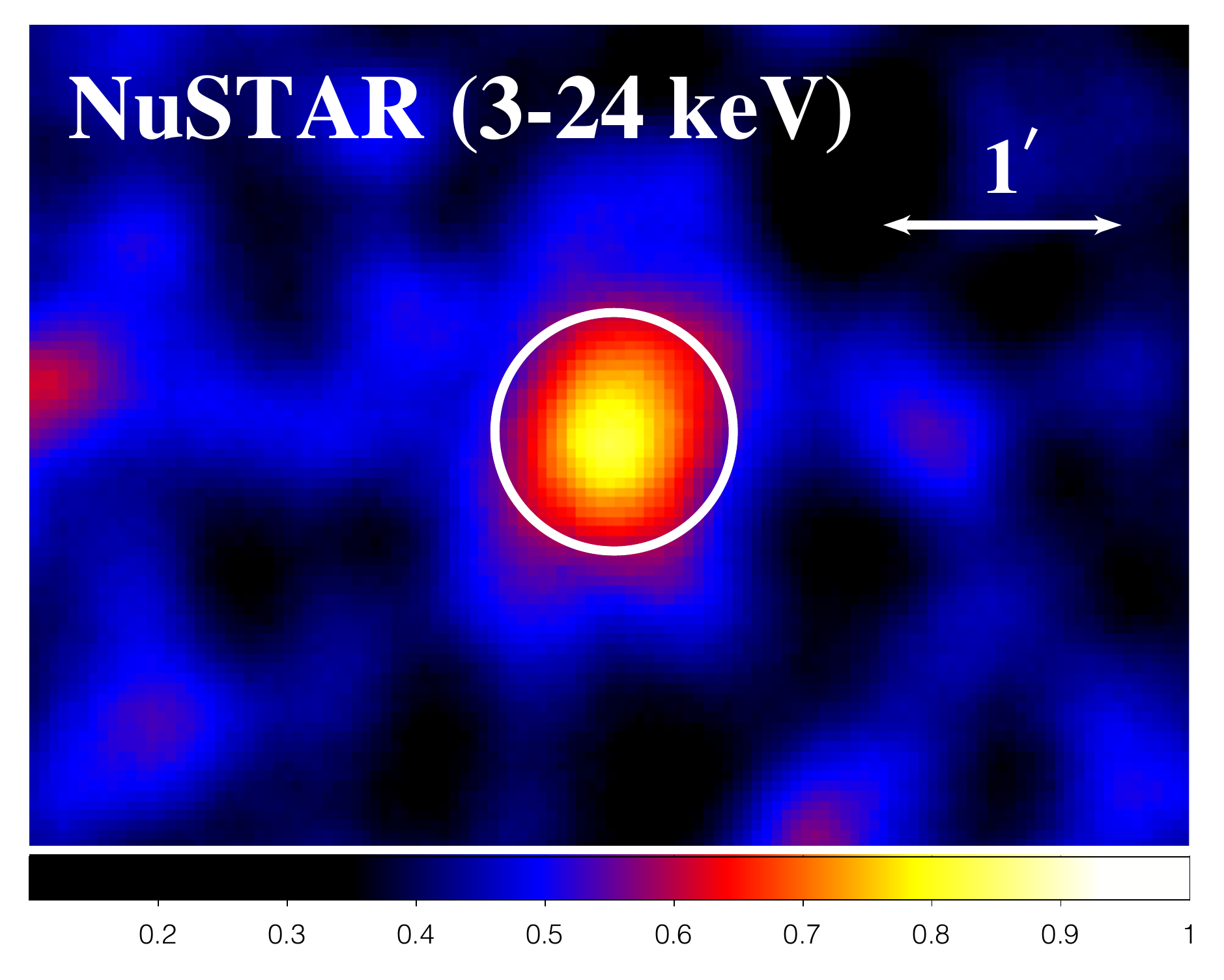}
    \caption{The 3--24 keV {\it NuSTAR} image (FPMA+FPMB) smoothed with a 2D Gaussian of a 1$\sigma$ radius of 5~pixels (12$\arcsec$.3).
    The white circle is centered on the peak of the emission and has a radius of 30$\arcsec$.}
    \label{fig:nustar-image}
\end{figure}

In the first observation, {\it NuSTAR} failed to detect significant signal from the source, with a 3$\sigma$ upper limit of 0.002 cts s$^{-1}$ in the 3--24 keV band, most probably owing to the limited exposure. 
In the second observation, the source was detected with a net count rate of (7.2 $\pm$ 1.1) $\times 10^{-4}$ cts s$^{-1}$ in the 3--24 keV band; a smoothed image around the target is displayed in Figure~\ref{fig:nustar-image}. 
This is a first detection by {\it NuSTAR} in terms of IR-bright DOG.
To avoid uncertainties due to possible variability between the two {\it NuSTAR} observation epochs, we decided not to utilize the data of the first observation in the following analysis. 
The 3--24 keV light curves in the second observation show no evidence for significant time variability on a time scale of 5820 sec. 
The spectra were binned to a minimum of more than 20 counts per energy bin in order to facilitate the use of $\chi^{2}$-statistics.

\subsection{XMM-Newton}

{\it XMM-Newton} \citep{Jansen2001} serendipitously observed WISE0825+3002 (ObsID: 0504102001) for a net exposure of 19.0~ks on 2007 November 3 with the EPIC/pn \citep{Struder2001} and EPIC/MOS \citep{Turner2001} cameras. 
We did not analyze the MOS data because of their low photon statistics. 
The data were reduced in a standard manner by using the {\it XMM-Newton} Science Analysis System \citep[\textsc{sas:}][]{Gabriel} v17.0.0 and Current Calibration Files (CCF) as of 2018 June 22.  
To produce calibrated event files, we used the \textsc{epproc} task. 
Since no background flare was observed in the light curve of PATTERN=0 events in the 10--12 keV band, we did not apply any time filter. 
The spectrum was extracted from a circular region of 30$\arcsec$ radius around the target, and the background was taken from a nearby source-free circular region with a radius of 60$\arcsec$. 
Only single and double pattern events (PATTERN 0--4) were used. 
The spectrum was binned to have a minimum counts of 20 per energy bin. 
The redistribution matrix file (RMF) and auxiliary response file (ARF) were generated with the \textsc{rmfgen} and
\textsc{arfgen} tasks, respectively.

\section{Results and discussions}
\label{result}

\subsection{X-ray luminosity and hydrogen column density}
\label{sec:x-ray-ana}

We jointly analyze the X-ray spectra obtained with {\it NuSTAR} in 2019 and with {\it XMM-Newton} in 2007, which covers the 3.0--30 keV and 0.35--8 keV bands with sufficient signal-to-noise ratios, respectively.
For spectral analysis, we utilize the \textsc{xspec} v.12.10.1 \citep{Arnaud1996} package, adopting the $\chi^2$ minimization algorithm. 
Galactic absorption of $N_{\rm H}^{\rm Gal} = 3.55 \times 10^{20}$ cm$^{-2}$ \citep{HI4PI2016}, modeled by
\textsf{phabs}, is always included in spectral fits. 
We assume the solar abundances by \citet{Anders1989} and the redshift $z$ = 0.89 (in Table~\ref{tab:analysis} we also show the maximum errors within the uncertainty of the photometric redshift). 
We ignore possible time variability between the two epochs (2007 and 2019), since it is not significantly required from the data.

\begin{deluxetable*}{llll}
\tablewidth{\textwidth}
\tablecaption{Summary of the X-ray Spectral Analysis of the Target \label{tab:analysis}}
\tablehead{
\colhead{Parameter} &
\colhead{Basic Model\ \ }   &
\colhead{Pexmon Model\ \ }   &
\colhead{XCLUMPY Model\ \ } 
}
\startdata
Column density ($N_{\rm H}$) [$10^{22}$ cm$^{-2}$]  &$95^{+70}_{-40}$ \,\,\,($^{+116}_{-46}$)  &$\geqslant28^{a}$ \,\,\,($\geqslant26^{a}$) &$99^{+82}_{-42}$ \,\,\,($^{+153}_{-49}$) \\
Scattering fraction ($f_{\rm scat}$) [\%]  &$2.4^{+2.7}_{-1.3}$ ($^{+2.7}_{-1.6}$)  &$9.2^{+8.9}_{-6.1}$ ($^{+11.9}_{-6.1}$)  &$3.9^{+3.3}_{-1.9}$ ($^{+3.4}_{-2.0}$) \\
Observed 2--10 keV flux ($F^{\rm obs}_{2-10}$) [10$^{-14}$ erg s$^{-1}$ cm$^{-2}$]  &3.5  &3.4  &3.5\\
Intrinsic 2--10 keV luminosity ($L_{\rm 2-10}$) [$10^{44}$ erg s$^{-1}$]  &$7.0^{+8.9}_{-3.4}$ ($^{+17.4}_{-4.9}$)  &$1.4^{+0.8}_{-0.5}$ ($^{+1.2}_{-0.7}$)  &$4.2^{+2.8}_{-1.6}$ ($^{+7.9}_{-2.7}$) \\
$\chi^2$/dof & 4.3/12 & 3.9/12 & 3.3/12 
\enddata
\tablecomments{The errors outside the parentheses correspond to the statistical errors at 90\% confidence limits. Those inside the parentheses denote the maximum intervals when the uncertainty in the photometric redshift is taken into account.}
\tablenotetext{a}{The column density reaches an upper limit $10^{25}$ cm$^{-2}$ allowed in the fits.}
\end{deluxetable*}

\subsubsection{Basic Model}

We first fit the observed spectra with a basic model that consists of a transmitted component through a cold absorber and a scattered component by surrounding gas.
In the \textsc{xspec} terminology, it is described as 
\begin{align}
&\mathsf{phabs * (zphabs * cabs * zpowerlw * zhighect} \notag\\
&\mathsf{+ const * zpowerlw * zhighect)}.
\end{align}
The intrinsic spectrum is modeled by a power law with a high-energy
exponential cutoff (\textsf{zpowerlw*zhighect}).
In this paper, we always  fix the photon index ($\Gamma$) at 1.8 and the high energy cutoff at 360 keV as typical values of AGNs
\citep[e.g.,][]{Ueda2014,Ricci2017c,Tanimoto2018}, which are difficult to  constrain from our data due to the limited photon statistics.
In the first term, we consider Compton scattering out of the line of sight (\textsf{cabs}), whose column density ($N_{\rm H}$) is linked to that of photometric absorption (\textsf{zphabs}).
The \textsf{const} factor in the second term represents the scattering fraction, $f_{\rm scat}$\footnote{Possible contribution from optically thin thermal emission and high-mass X-ray binaries (HMXBs) in the host galaxy may be included in this component.} \citep[e.g.,][]{Ueda2007}.
We define it as the ratio of the unabsorbed fluxes at 1 keV between the primary and scattered components, whose normalizations are tied together.
This model reproduces the observed spectra well ($\chi^{2}$/dof = 4.3/12).
Table~\ref{tab:analysis} lists the best-fit parameters, together  with an intrinsic luminosity in the rest-frame 2--10 keV band. 
We obtain a line-of-sight column density of  $N_{\rm H}$ = $1.0^{+0.7}_{-0.4} \times 10^{24}$~cm$^{-2}$.
The spectra unfolded with the energy responses and the best-fit model are plotted in Figure~\ref{fig:spectra}.

\subsubsection{Pexmon Model}

\begin{figure*}
    \epsscale{1.12}
    \plottwo{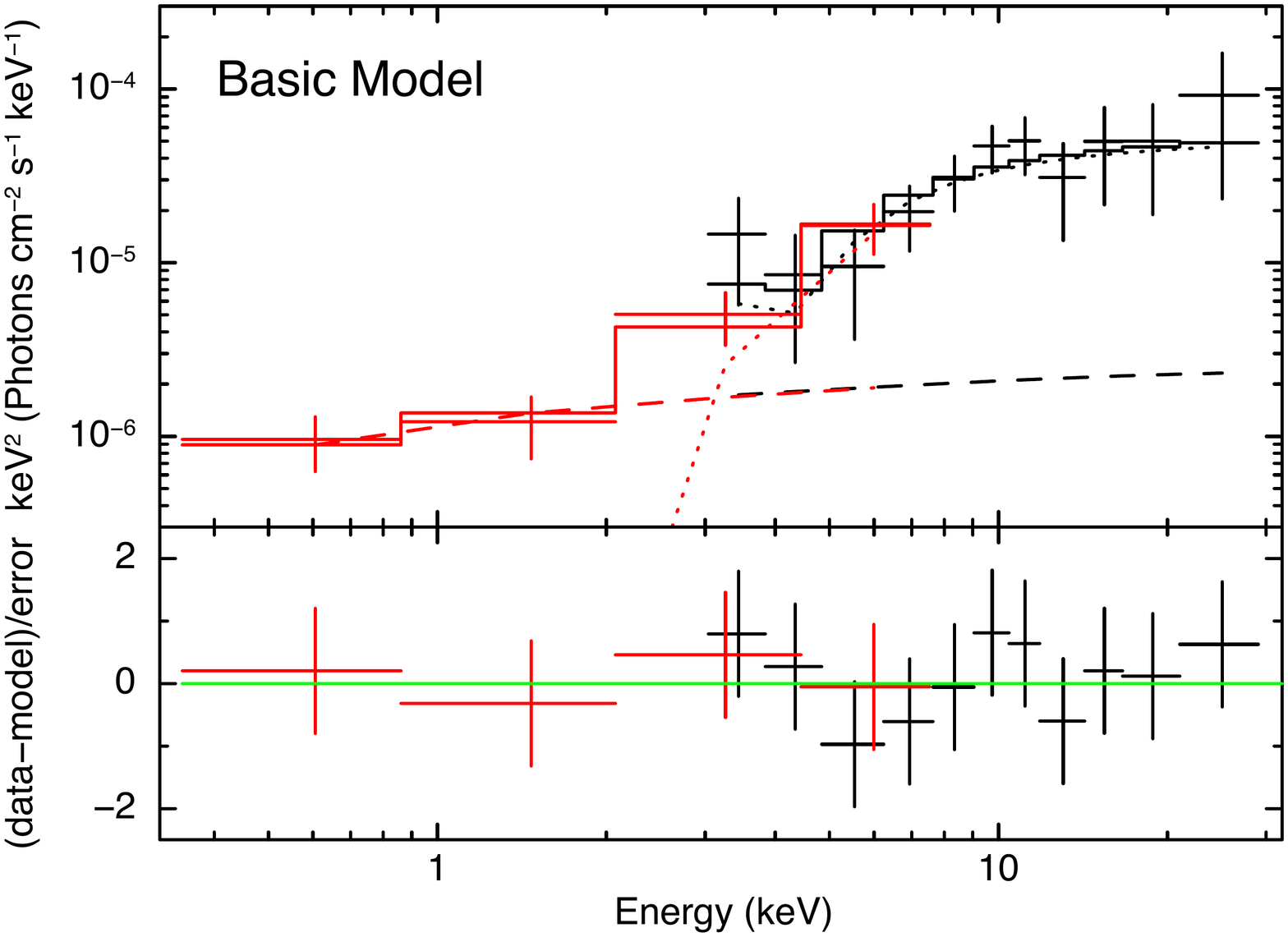}{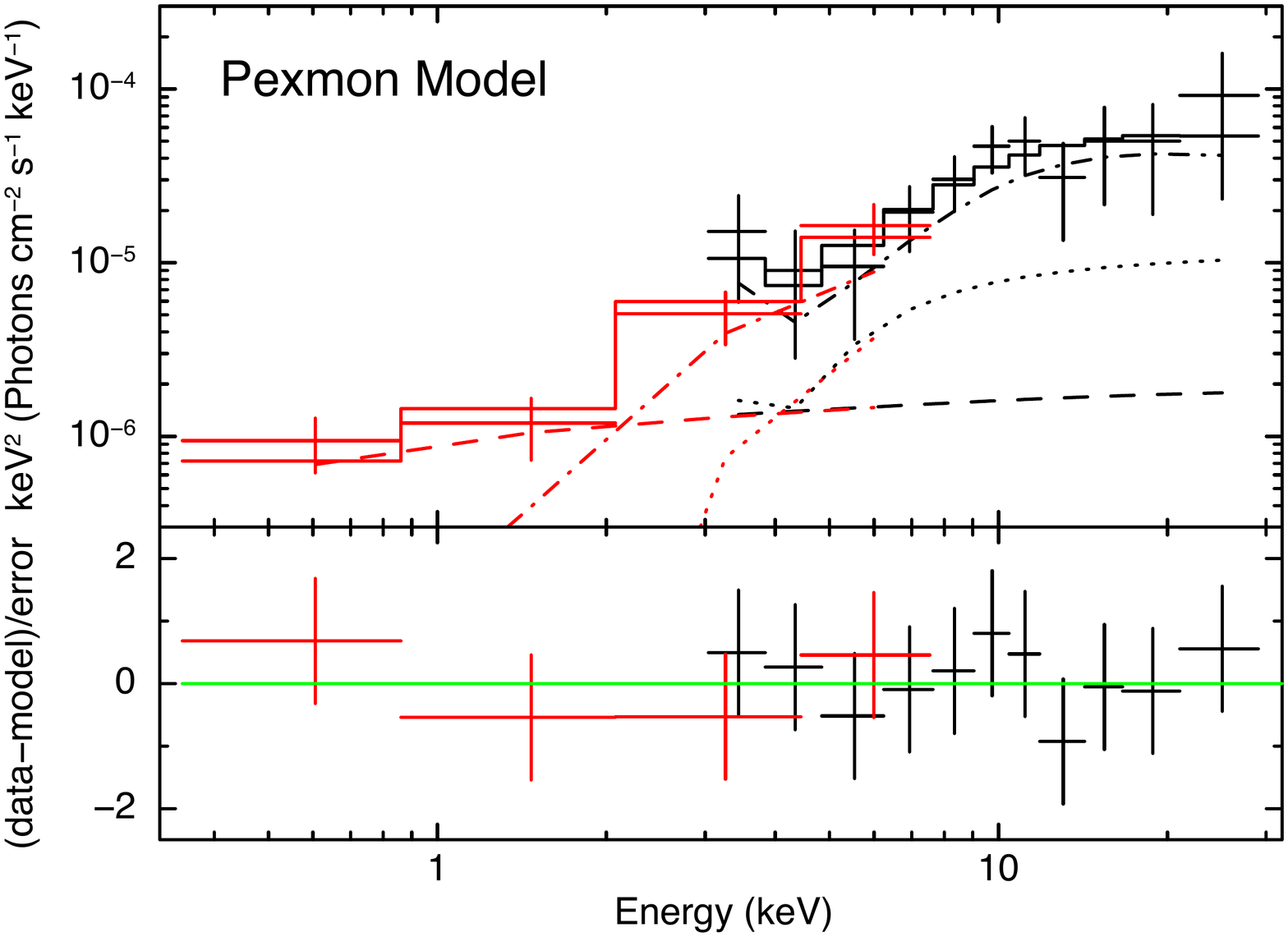}
    \includegraphics[keepaspectratio,scale=0.32]{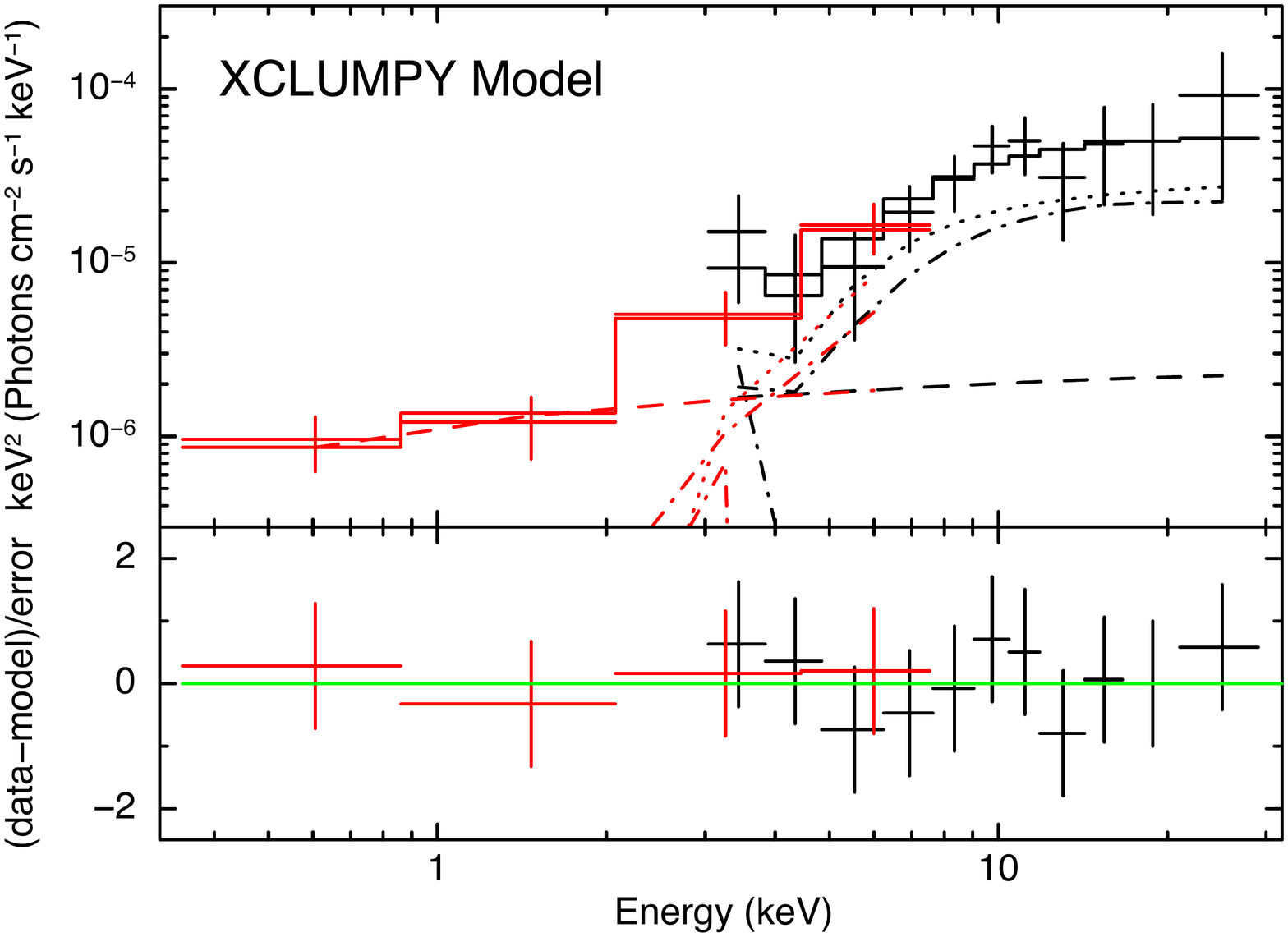}
    \caption{The unfolded {\it NuSTAR}/FPM (black) and {\it XMM-Newton} EPIC/pn (red) spectra of WISE0825+3002 
fit with the basic model (top left), pexmon model (top right), and XCLUMPY model (bottom).
    The solid, dotted, dashed, and dotted-dashed lines correspond to the total, cutoff power-law component, scattered component, and reflection component (e.g., reflection continuum and Fe K$\alpha$ emission line), respectively.
    The bottom panels show the residuals.
    }
    \label{fig:spectra}
\end{figure*}

As a more realistic model, we next add a reflection component from  surrounding cold material, which is known to be commonly present in obscured AGNs \citep[e.g.,][]{Turner1997,Kawamuro2016}. 
In the \textsc{xspec} terminology, the model is expressed as 
\begin{align}
&\mathsf{phabs * (zphabs * cabs * zpowerlw * zhighect} \notag\\
&\mathsf{+ const * zpowerlw * zhighect + pexmon)}. 
\end{align}
The first and second terms are the same as in the previous model.
The third term approximately represents a reflection component from cold matter in the circumnuclear region.
Here we adopt the \textsf{pexmon} code \citep{Nandra2007}, which calculate a reflected continuum along with Fe and Ni K fluorescence lines.
The photon index and power-law normalization are linked to those of the primary component.
The reflection strength, defined as $R = \Omega/2\pi$ ($\Omega$ is the solid angle of the reflector), is fixed at $R=1$.
The inclination angle is set to 60$^\circ$ as a representative value.
The model is also found to well reproduce the spectra ($\chi^{2}$/dof = 3.9/12), yielding a line-of-sight column density of  $N_{\rm H}$ $\geqslant 2.8 \times 10^{23}$~cm$^{-2}$. 
The best-fit parameters are summarized in Table~\ref{tab:analysis}, and the best-fit model is plotted in Figure~\ref{fig:spectra}.
If we instead assume $R = 1.5$ or $R = 0.5$, we obtain $N_{\rm H}$ $\geqslant 2.2 \times 10^{23}$~cm$^{-2}$ or $N_{\rm H}$ $\geqslant 3.8 \times 10^{23}$~cm$^{-2}$, respectively.

\subsubsection{XCLUMPY Model}
\label{S_X}

We finally apply the XCLUMPY model \citep{Tanimoto2019}, a numerical spectral model from clumpy tori in AGNs.
Since there are many pieces of evidence suggesting that AGN tori are not smooth but have clumpy structure (see \citealt{Tanimoto2019} for details), we regard this model as the most realistic one compared with the previous two models.
XCLUMPY reproduces the reflection component from a clumpy torus whose geometry is defined in the same way as in the
CLUMPY model \citep{Nenkova2008a,Nenkova2008b}, which has been used for infrared studies.
The torus parameters are the column density along the equatorial plane ($N_{\rm H}^{\rm Equ}$), the torus angular width ($\sigma$), and the inclination angle ($i$).\footnote{The other parameters, the inner and outer radii of the torus (0.05 pc and 1.00 pc), the radius of each clump (0.002 pc), number of the clump along the equatorial plane (10.0), and the index of radial density profile (0.5), are fixed \citep{Tanimoto2019}.} 
In \textsc{xspec} terminology, the model is expressed as
\begin{align}
&\mathsf{phabs * (zphabs * cabs * zpowerlw * zhighect} \notag\\
&\mathsf{+ const * zpowerlw * zhighect} \notag\\
&\mathsf{+ atable\{xclumpy\_R.fits\} + atable\{xclumpy\_L.fits\})}. 
\end{align}
The first (transmitted component) and second (scattered component) terms are the same as in the previous models.
The third and fourth ones correspond to the two table models of XCLUMPY, the reflection continuum and fluorescence emission lines, respectively.
The torus angular width and the inclination angle are fixed at 30$^\circ$ and 60$^\circ$, respectively, which cannot be constrained  from our data; we have confirmed that the choice of these parameters does not significantly affect our results.
We find that this model also gives a good fit ($\chi^{2}$/dof = 3.3/12).
This fit is statistically better compared to the other models, supporting that this model is a physically more realistic description of the spectrum.
The best-fit parameters are summarized in Table~\ref{tab:analysis}, and the best-fit model is plotted in Figure~\ref{fig:spectra}.
For a given torus geometry, we can convert the equatorial hydrogen column density into the line-of-sight one ($N_{\rm H}$) by Equation (3) in \citet{Tanimoto2019}.
We find that this galaxy contains a mildly CT AGN with a line-of-sight absorption of 
$N_{\rm H}$ = $1.0^{+0.8}_{-0.4} \times 10^{24}$ cm$^{-2}$.
The rest-frame 2--10 keV intrinsic luminosity obtained is $4.2^{+2.8}_{-1.6} \times 10^{44}$ erg~s$^{-1}$.
In the following discussion, we adopt these values as the most  reliable estimates of the column density and intrinsic luminosity.\\

\begin{figure}
\plotone{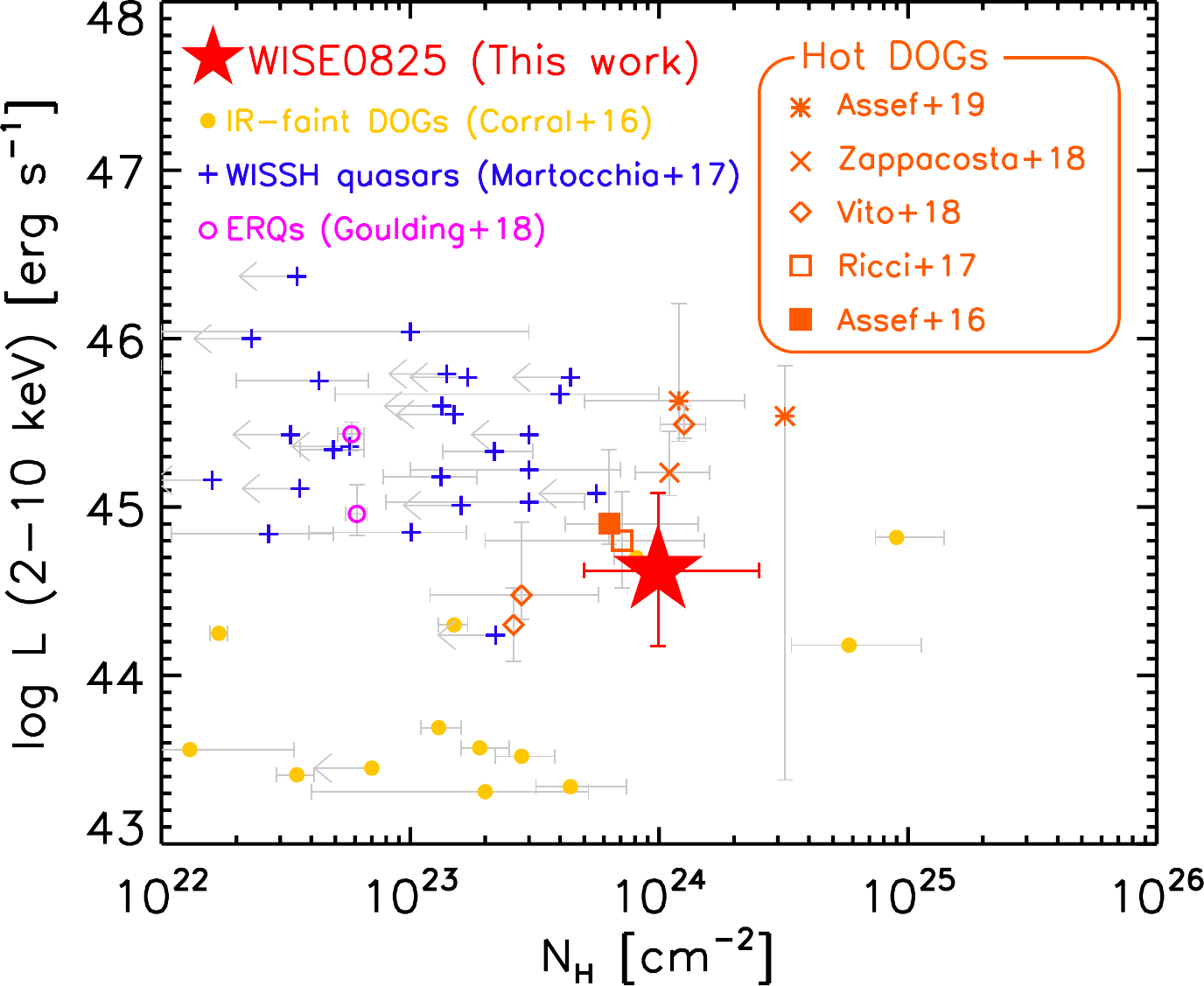}
\caption{Hard X-ray luminosity in the rest-frame 2--10 keV band as a function of hydrogen column density ($N_{\rm H}$). Yellow circles, blue crosses, and magenta circles represent IR-faint DOGs \citep{Corral}, WISSH quasars \citep{Martocchia}, and ERQs \citep{Goulding}, respectively. Orange symbols represent hot DOGs \citep{Assef,Ricci,Vito,Zappacosta,Assef_19}. Red star represents WISE0825+3002.}
\label{NH}
\end{figure}

Figure \ref{NH} shows the absorption-corrected hard X-ray luminosity in the rest-frame 2--10 keV band, $L_{\rm X}$ (2--10 keV), and $N_{\rm H}$ of WISE0825+3002, where uncertantiy of its redshift is taken into account of error bar.
IR-faint DOGs with flux density at 24 $\micron$ $<$ 1.0 mJy detected in {\it Chandra} deep field \citep{Georgantopoulos,Corral}, extremely red quasars \citep[ERQs:][]{Ross,Hamann,Goulding}, and hyper-luminous quasars, selected from the SDSS and {\it WISE} \citep[WISSH quasars:][]{Bischetti,Martocchia} are also plotted.
We also plotted $L_{\rm X}$ (2--10 keV) and $N_{\rm H}$ for hot DOGs \citep{Assef,Ricci,Vito,Zappacosta,Assef_19}.

We found that the distribution of WISSH quasars and ERQs in $N_{\rm H}-L_{\rm X}$ plane is different from that of (hot) DOGs, as reported by \cite{Vito}.
Among DOG population, WISE0825+3002 (IR-bright DOG) may be located between IR-faint DOGs and hot DOGs in $N_{\rm H}-L_{\rm X}$ plane.
Given a same $N_{\rm H}$, X-ray luminosity of WISE0825+3002 is smaller than that of hot DOGs, which suggests that accreting power of IR-bright DOGs is moderate compared with hot DOGs (see also Section \ref{S_BH}).

\subsection{Host properties derived from the SED fitting}
\label{S_SED}

In order to derive physical properties of WISE0825+3002 such as stellar mass and SFR, we carried out the SED fitting with the code investigating galaxy emission \citep[{\sf CIGALE};][]{Burgarella,Noll,Ciesla_15,Ciesla_16,Boquien} conducting a SED modeling with stellar, AGN, and SF components by taking into account the energy balance between the absorbed energy emitted in UV/optical from SF/AGN and the re-emitted energy in IR from dust.
Input parameters are basically same as what \cite{Toba_19a} adopted.
We applied a delayed star formation history (SFH) assuming a single starburst with an exponential decay.
For single stellar population (SSP) and attenuation low, we adopted the stellar templates of \cite{Bruzual} with \cite{Calzetti} dust extinction low assuming \cite{Chabrier} initial mass function.
We also added the standard default nebular emission model \citep{Inoue}.
AGN emission is modeled by an AGN model provided by \cite{Fritz} while  dust emission is modeled by dust templates of \cite{Dale} \citep[see also][]{Matsuoka,Toba_19b}.

\begin{figure}
\plotone{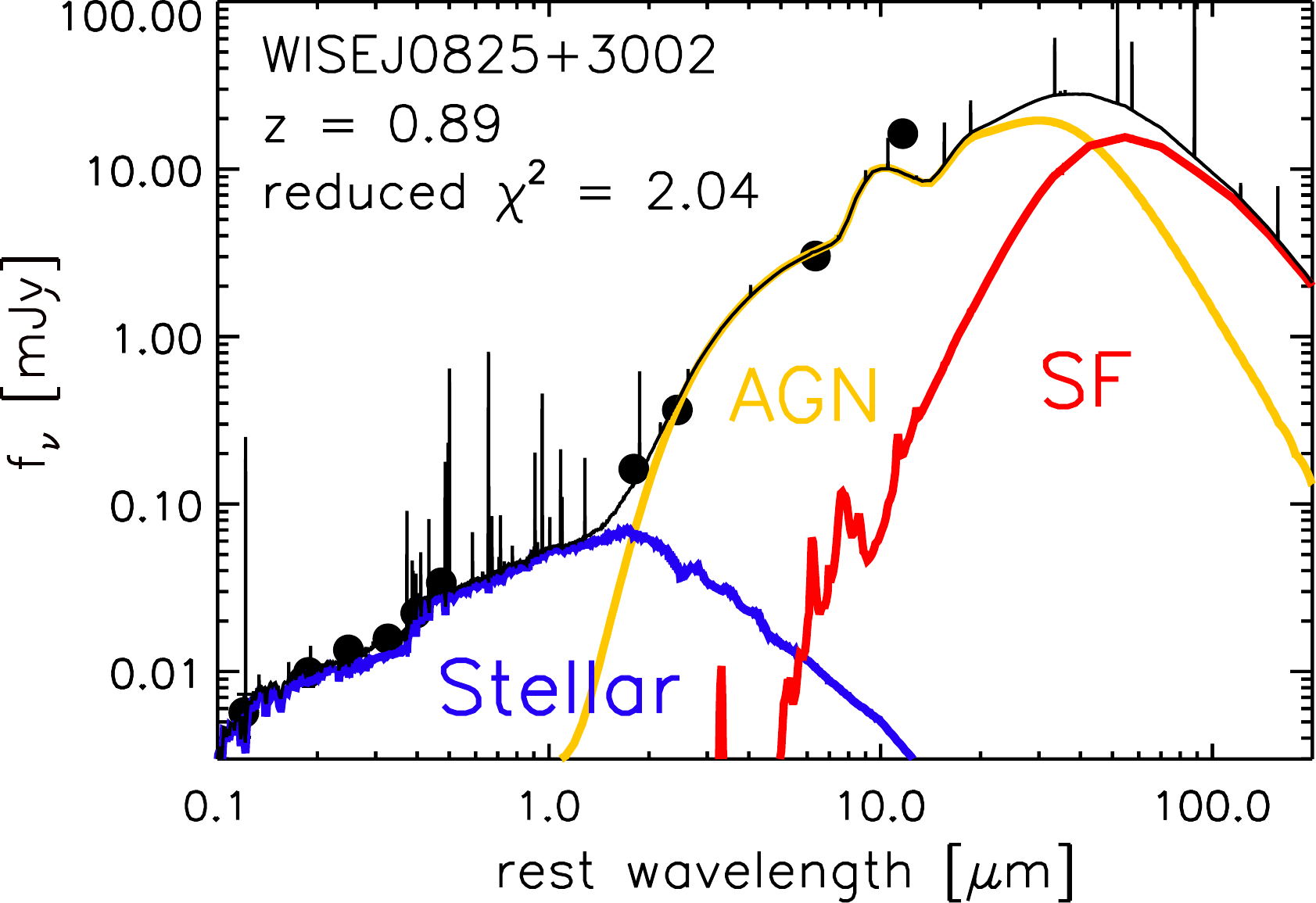}
\caption{SED of WISE0825+3002. The back points are photometric data. The contribution from the stellar, AGN, and SF components to the total SED are shown as blue, yellow, and red line, respectively. The black solid line represents the resultant SED.}
\label{SED}
\end{figure}

Figure \ref{SED} shows the result of the SED fitting.
The observed data points of WISE0825+3002 are well-fitted by the combination of stellar and AGN components with a moderately good reduced $\chi^{2}$ (= 2.04) although SF component may not be constrained well due to the lack of far-IR (FIR) data.
The physical properties derived by {\sf CIGALE} are summarized in Table \ref{Table}. 
The uncertainty of photometric redshift was also incorporated into the uncertainty of derived physical quantities that was estimated based on the Monte Carlo algorithm in the same manner as \cite{Toba_19b}.
The resultant color excess of stellar component ($E(B-V)_*$) is 0.21 $\pm$ 0.01.
The IR luminosity, $L_{\rm IR}$ (8--1000 $\micron$)\footnote{We integrated the best-fit SED over a wavelength range of 8--1000 $\micron$ to calculate the IR luminosity.}, is $(1.1 \pm 0.6) \times 10^{46}$ erg s$^{-1}$.
We found that AGN fraction, i.e., $L_{\rm IR}$ (AGN)/$L_{\rm IR}$ is $\sim$ 0.8 confirming that WISE0825+3002 is an  AGN-dominant object.
The derived stellar mass and SFR are $(5.3 \pm 4.4) \times 10^{10}$ $M_{\sun}$ and 85 $\pm$ 39 $M_{\sun}$ yr$^{-1}$, respectively, where SFR was estimated based only on resultant parameters of SFH output by {\tt CIGALE} \citep[see][for more detail]{Boquien}.
This means that WISE0825+3002 lies above the main sequence of normal star-forming galaxies at similar redshift on $M_{*}$--SFR plane \citep[e.g.,][]{Elbaz,Pearson}, indicating that WISE0825+3002 has an active star formation.
This trend is roughly consistent with that of other IR-bright DOGs \citep{Toba_17b}.

\subsection{Black hole properties}
\label{S_BH}
Finally, we discuss the BH properties such as BH mass ($M_{\rm BH}$) and Eddington ratio ($\lambda_{\rm Edd}$) of WISE0825+3002.
The BH mass was estimated from stellar mass by using an empirical relation with a scatter of 0.28 dex, reported in \cite{Kormendy}, and we then converted it to Eddington luminosity ($L_{\rm Edd}$). 
The bolometric luminosity ($L_{\rm bol}$) was derived by integrating the best-fit SED template of AGN component output by {\sf CIGALE} over wavelengths longward of Ly$\alpha$ \citep[see][for more detail]{Toba_17c}.
We note that the expected 2--10 keV X-ray bolometric correction, $\kappa_{\rm 2-10\,\,keV} = L_{\rm bol}/L_{\rm X}$ (2--10 keV), is about 38 that is in good agreement with what reported by \cite{Vasudevan} \citep[see also][]{Ricci_16,Yamada}.

\begin{figure}
\plotone{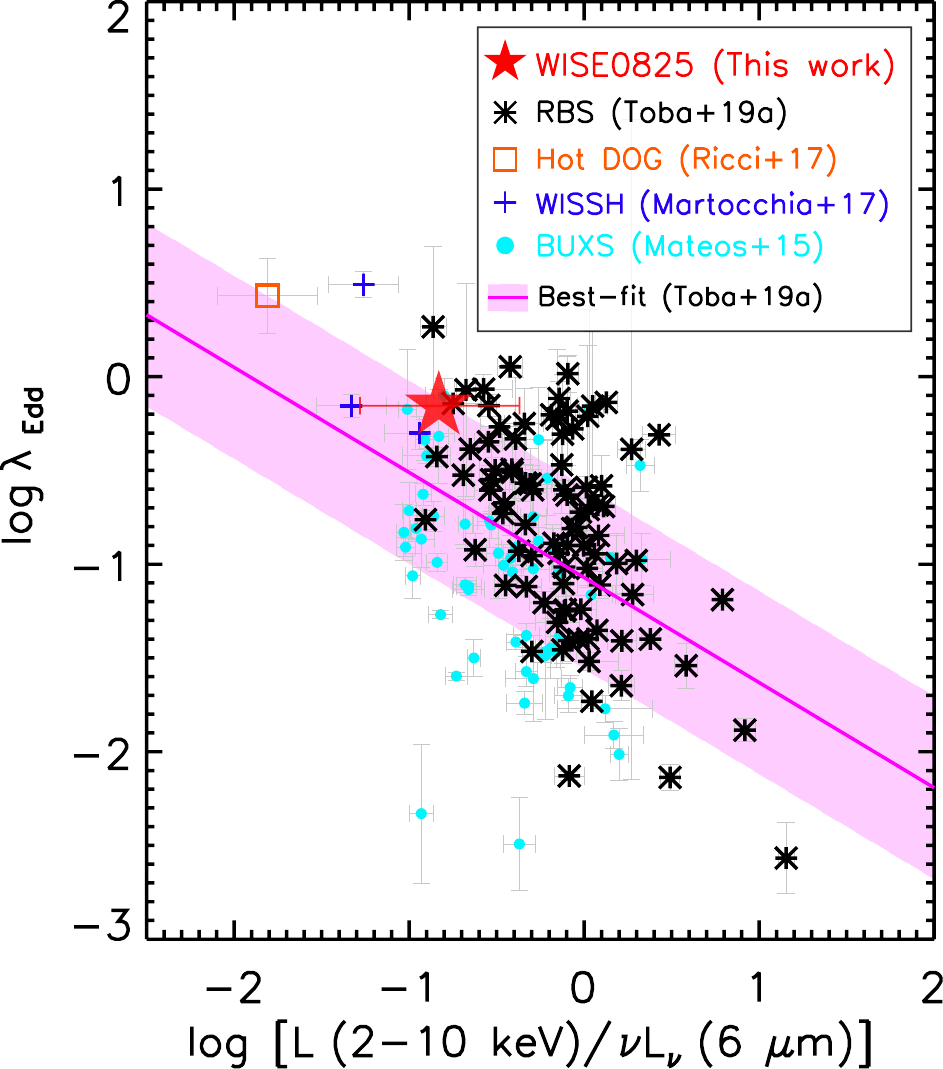}
\caption{Eddington ratio ($\lambda_{\rm Edd}$) as a function of $L_{\rm X}/L_{\rm 6}$ of RBS type 1 AGNs \citep[black asterisk: ][]{Toba_19a}, BUXS type 1 AGNs \citep[cyan circle: ][]{Mateos}, a hot DOG \citep[orange square: ][]{Ricci}, and WISSH quasars \citep[blue crosses: ][]{Bischetti,Martocchia}. Red star represents WISE0825+3002. A magenta solid line with shaded region is a linear relation between $\lambda_{\rm Edd}$ and $L_{\rm X}/L_{\rm 6}$ reported by \cite{Toba_19a}.
\label{Edd}}
\end{figure}

The resultant BH mass and Eddington ratio ($\lambda_{\rm Edd} = L_{\rm bol}/L_{\rm Edd}$) are $2.5 \times 10^{8}$ $M_{\sun}$ and 0.70, respectively. 
Figure \ref{Edd} shows the Eddington ratio as a function of the luminosity ratio of hard X-ray in the 2--10 keV band and 6 $\micron$ ($L_{\rm X}/L_{\rm 6}$) that error is taken into account the unceranty of redshift.
The X-ray luminosity was corrected for the absorption (see Section \ref{S_X}) while 6 $\micron$ luminosity was corrected for the  contamination of the host galaxy in the same manner as \cite{Toba_19a}.
In this figure, we plot type 1 AGNs \citep{Toba_19a} selected by using {\it ROSAT} Bright Survey (RBS) catalog \citep{Fischer,Schwope} and type 1 AGNs drawn from the Bright Ultra-hard {\it XMM-Newton} Survey \citep[BUXS;][]{Mateos}.
A hot DOG \citep{Ricci} and WISSH quasars \citep{Martocchia} are also plotted \footnote{If uncertainty of $\lambda_{\rm Edd}$, $L_{6}$, or $L_{X}$, was not provided, we conservatively assumed 20 per cent error of the corresponding quantity \citep[see][for details]{Toba_19a}.}.
The BH masses of BUXS type 1 AGNs, the hot DOG, and WISSH quasars are estimated from broad emission lines such as Mg{\,\sc ii} and H$\beta$.
\cite{Toba_19a} reported that there is a negative correlation between the $\lambda_{\rm Edd}$ and $L_{\rm X}/L_{\rm 6}$ suggesting that AGNs with high Eddington ratio (i.e., with high accretion efficiency) tend to show the X-ray deficit compared to MIR emission. 
We found that WISE0825+3002 also follows this correlation.

The relation between $N_{\rm H}$ and $\lambda_{\rm Edd}$ of WISE0825+3002 suggests that this object may correspond to a blow-out phase \citep{Fabian_08,Fabian_09,Ricci2017d}.
Indeed, a large fraction of IR-bright DOGs show a strong ionized gas outflow \citep[][see also \citealt{Noboriguchi}]{Toba_17c}, supporting the above possibility.

\section{Summary and Conclusions}
\label{Conc}

In this paper, we report the discovery of a CT AGN, WISE J082501.48+300257.2 (WISE0825+3002) at $z_{\rm photo} = 0.89 \pm 0.18$.
By performing hard X-ray observations with {\it NuSTAR} and spectral analysis with XCLUMPY model, we estimate the hard X-ray luminosity in the rest-frame 2--10 keV band and hydrogen column density of WISE0825+3002 to be $4.2^{+2.8}_{-1.6} \times 10^{44}$ erg s$^{-1}$ and $1.0^{+0.8}_{-0.4} \times 10^{24}$ cm$^{-2}$, respectively, making it mildly CT AGN.

We also conduct the SED fitting with {\tt CIGALE} to investigate host properties.
The resultant stellar mass, SFR, and IR luminosity are  $ (5.3 \pm 4.4) \times 10^{10}$ $M_{\sun}$,  85 $\pm$ 39 $M_{\sun}$ yr$^{-1}$, and $(1.1 \pm 0.6) \times 10^{46}$ erg s$^{-1}$, respectively.
The BH mass converted from the stellar mass by using an empirical relation and the Eddington ratio are $2.5 \times 10^{8}$ $M_{\sun}$ and  0.70, respectively. 
The relation between luminosity ratio of hard X-ray and MIR, and Eddington ratio of WISE0825+3002 follows a correlation \cite{Toba_19a} reported.

According to the fact that (i) {\it WISE} W1 (3.4 $\micron$) and W2 (4.6 $\micron$) color of hot DOGs is redder than that of IR-bright DOGs and W1--W2 color is correlated to the AGN activity \citep{Blecha}, (ii) $L_{\rm X}$ (2--10 keV) of hot DOGs is larger than that of IR-bright DOGs with a similar $N_{\rm H}$ (Figure \ref{NH}), (iii) $\lambda_{\rm Edd}$ of hot DOGs seems to be larger than that of IR-bright DOGs (Figure \ref{Edd}), and (iv) the number density of hot DOGs is much smaller than that of IR-bright DOGs \citep{Assef_15,Toba_15}, hot DOGs are more specific and short-lived phase in which SMBH is actively growing, compared with IR-bright DOGs.
This indicates that a comprehensive work on hot DOGs and IR-bright DOGs is crucial to investigate an evolutionary link between two population and to understand the growth history of SMBHs.

\acknowledgments
We gratefully acknowledge the anonymous referee for a careful reading of the manuscript and very helpful comments.
We also thank Prof. Denis Burgarella for helping us to understand {\tt CIGALE} code.

This work makes use of data from the {\it NuSTAR} mission, a project led by Caltech, managed by the Jet Propulsion Laboratory, and funded by NASA. 
We thank the {\it NuSTAR} Operations, Software, and Calibration teams for their support with the execution and analysis of these observations. This research has made use of the {\it NuSTAR} Data Analysis Software, jointly developed by the ASI Science Data Center (Italy) and Caltech.

This research has made use of data and/or software provided by the High Energy Astrophysics Science Archive Research Center (HEASARC), which is a service of the Astrophysics Science Division at NASA/GSFC and the High Energy Astrophysics Division of the Smithsonian Astrophysical Observatory.

This work is based on archival data from the {\it Galaxy Evolution Explorer} which is operated for NASA by the California Institute of Technology under NASA contract NAS5-98034.

Funding for SDSS-III has been provided by the Alfred P. Sloan Foundation, the Participating Institutions, the National Science Foundation, and the U.S. Department of Energy Office of Science. The SDSS-III web site is http://www.sdss3.org/.
SDSS-III is managed by the Astrophysical Research Consortium for the Participating Institutions of the SDSS-III Collaboration including the University of Arizona, the Brazilian Participation Group, Brookhaven National Laboratory, Carnegie Mellon University, University of Florida, the French Participation Group, the German Participation Group, Harvard University, the Instituto de Astrofisica de Canarias, the Michigan State/Notre Dame/JINA Participation Group, Johns Hopkins University, Lawrence Berkeley National Laboratory, Max Planck Institute for Astrophysics, Max Planck Institute for Extraterrestrial Physics, New Mexico State University, New York University, Ohio State University, Pennsylvania State University, University of Portsmouth, Princeton University, the Spanish Participation Group, University of Tokyo, University of Utah, Vanderbilt University, University of Virginia, University of Washington, and Yale University.

This publication makes use of data products from the {\it Wide-field Infrared Survey Explorer}, which is a joint project of the University of California, Los Angeles, and the Jet Propulsion Laboratory/California Institute of Technology, funded by the National Aeronautics and Space Administration.

This work is supported by JSPS KAKENHI Grant numbers 18J01050 and 19K14759 (Y.Toba), 19J22216 (S.Yamada), 17K05384 (Y.Ueda), 16H03958, 17H01114, and 19H00697 (T.Nagao), 16K05296 (Y.Terashima), and 17J06407 (A.Tanimoto).
Y.Toba and W.H.Wang acknowledge the support from the Ministry of Science and Technology of Taiwan (MOST 105-2112-M-001-029-MY3). 
C.Ricci acknowledges the CONICYT+PAI Convocatoria Nacional subvencion a instalacion en la academia convocatoria a\~{n}o 2017 PAI77170080.

\vspace{5mm}
\facilities{{\it NuSTAR}, {\it XMM-Newton}, {\it GALEX}, Sloan, {\it WISE}}


\software{IDL, IDL Astronomy User's Library \citep{Landsman}, {\sf XCLUMPY} \citep{Tanimoto2019}, {\sf HEAsoft} 6.25, {\sf XSPEC} \citep{Arnaud1996}, {\sf SAS} 17.00 \citep{Gabriel}, {\sf CIGALE} \citep{Boquien}}


\end{document}